\documentclass[prl,twocolumn,superscriptaddress,amsmath]{revtex4}
\usepackage{graphicx}

\begin{document}

\def\ket#1{|#1\rangle} 
\def\bra#1{\langle#1|}
\def\av#1{\langle#1\rangle}
\def\myarrow{\mathop{\longrightarrow}}

\title{Entangled and disentangled evolution for a single atom in a driven cavity}

\affiliation{Department of Physics, University of Arkansas, Fayetteville, AR 72701\\}\affiliation{Raytheon Company, Science and Technology, Garland, TX 75042\\}\affiliation{Department of Physics, Miami University, Oxford, OH 45056\\}\affiliation{Department of Physics, University of Maryland, College Park, MD 20742\\}

\author{J. Gea-Banacloche}
\affiliation{Department of Physics, University of Arkansas, Fayetteville, AR 72701\\}
\affiliation{Department of Physics, University of Maryland, College Park, MD 20742\\}%
\author{T. C. Burt}
\affiliation{Raytheon Company, Science and Technology, Garland, TX 75042\\}
\author{P. R. Rice}
\affiliation{Department of Physics, Miami University, Oxford, OH 45056\\}\affiliation{Department of Physics, University of Maryland, College Park, MD 20742\\}
\author{L. Orozco}
\affiliation{Department of Physics, University of Maryland, College Park, MD 20742\\}

\date{\today}

\begin{abstract}
For an atom in an externally driven cavity, we show that special initial states lead to near-disentangled atom-field evolution, and superpositions of these can lead to near maximally-entangled states.  Somewhat counterintutively, we find that (moderate) spontaneous emission in this system actually leads to a transient increase in entanglement beyond the steady-state value.  We also show that a particular field correlation function could be used, in an experimental setting, to track the time evolution of this entanglement.      \end{abstract}
\maketitle

In a recent, beautiful experiment, Auffeves and coworkers \cite{auffeves} have verified the prediction \cite{myself} that, for a two-level atom interacting with a single mode of the electromagnetic field, in a coherent state with a moderately large number of photons, the natural evolution of the system leads to a superposition state in which two different states of the atom are correlated with two distinguishable states of the field, in what is often referred to as a ``Schr\"odinger cat.''   Such superpositions of correlated states, if pure, are naturally entangled states.  Interestingly, the possibility of preparing such entangled superpositions in the above system (which is described by the so-called Jaynes-Cummings model, or JCM) arises from the existence, in this model, of special trajectories along which the joint evolution of field and atom is to a good approximation {\em unentangled\/}, i.e., factorizable.  It is the coherent superposition of such trajectories that results in an entangled state.

The purpose of this paper is to show that a similar situation arises in another system of interest, which may be easier to prepare and observe in the optical domain, namely, a single atom in an externally driven cavity.  Unlike the JCM, this is an \emph{open} system, yet, as we shall show here, approximately factorizable trajectories exist in the absence of spontaneous emission.  Moreover, somewhat surprisingly, we find that the inclusion of spontaneous emission actually helps to create transient entangled states which are typically more entangled than the steady state.  We also show that there is a particular field correlation function that might be used to keep track, in ``real time,'' of the physical processes responsible for the evolution of this entangled state. (We note that there is a growing interest in exploring the connections between ``quantum optics''-style correlation functions and entangled states; see, e.g., \cite{wodk}, and references therein.)

The system we describe here has been studied before in great detail by, among others, Alsing and Carmichael \cite{paul} and Mabuchi and coworkers \cite{wiseman}.  The transient regime we are interested in here, however, has escaped attention in most of these previous studies, because they make a ``secular approximation'' on the Rabi frequency that results in an atom-field state that is explicitly disentangled at all times (as can be easily seen, for instance, from the ansatz in \cite{wiseman}).  An important exception is \cite{howard}, where the splitting of the field states in phase space, that plays an essential role in what follows, was explicitly discussed and illustrated, although the question of entanglement was not quantitatively addressed there. (For a very recent discussion of how different types of environment monitoring of this system may lead to different ways to characterize its entanglement, see also \cite{nha}.)

The starting point for our analysis is the following master equation for the joint atom-field density operator $\rho$:
\begin{align}
\frac{d}{dt}\rho = &-ig[a^\dagger\sigma_-+a\sigma_+,\rho] +{\cal E}[a^\dagger - a,\rho] \notag \\
&+ \kappa (2a\rho a^\dagger - a^\dagger a \rho - \rho a^\dagger a) \notag \\
&+\frac{\gamma}{2} (2\sigma_-\rho\sigma_+ - \sigma_+\sigma_-\rho - \rho\sigma_+\sigma_-)
\label{drhodt}
\end{align}
Here, $g$ is the atom-field coupling constant, for the cavity field mode, which is described by creation and annihilation operators $a^\dagger$ and $a$; $\sigma_-$ and $\sigma_+$ are the atom's raising and lowering operators; ${\cal E}$ is the amplitude of the external, driving field; $\kappa$ is the cavity loss rate, and $\gamma$ is the spontaneous emission rate.

As discussed in \cite{tcburt}, in the absence of spontaneous emission, approximately unentangled, quasi-pure state trajectories for this system are obtained whenever the initial joint atom-field state is of the form
\begin{equation}
\ket{\Psi^0_\pm(r_0,\phi_0)} = \frac{1}{\sqrt 2}\left(e^{-i\phi_0}\ket e \pm \ket g \right)\ket{r_0 e^{-i\phi_0}}
\label{psi0}
\end{equation}
where $\ket e$ and $\ket g$ are the atomic excited and ground states, respectively, and $\ket{r_0 e^{-i\phi_0}}$ is a field coherent state of arbitrary amplitude $r_0$ and phase $\phi_0$.  Trajectories starting from these special states remain approximately factorizable and quasi-pure for fairly long times, in spite of the dissipation represented by the term $\kappa$ in Eq.~(\ref{drhodt}); they retain approximately the same form as (\ref{psi0}), only with a time-dependent phase $\phi_\pm(t)$ for the field and the atomic dipole, and (in general) a time-dependent amplitude $r_\pm(t)$ for the field as well: 
\begin{equation}
\ket{\Psi_\pm(t;r_0,\phi_0)} = e^{i\Phi_{\pm}(t)} \ket{\Psi^0_\pm(r_\pm(t),\phi_\pm(t))}.
\label{psit}
\end{equation}
The overall phase $\Phi_{\pm}(t)$ will be discussed shortly below.  For $\phi_\pm(t)$ and $r_\pm(t)$, however, we note that consistency requires that they approximately obey the semiclassical equations of motion, derived from (\ref{drhodt}) by factoring the expectation values of atom-field operator products.  If one further treats the field as a classical quantity in these equations, one finds, for ${\cal E} > g/2$ (the so-called ``strong driving'' condition), a pair of steady states, with phases $\phi_{u,l}  = \mp\phi_{ss} = \mp \sin^{-1}(g/2{\cal E})$ and amplitude $r_{ss} = ({\cal E}/\kappa)\cos\phi_{ss}$.  (The subscripts $u$ and $l$ refer, respectively, to the ``upper'' and ``lower'' steady state, and follow the notation of \cite{paul}.)  The corresponding states of the quantum system are respectively $\ket{\Psi_u^0}\equiv\ket{\Psi^0_+(r_{ss},-\phi_{ss})}$ and $\ket{\Psi_l^0}\equiv\ket{\Psi^0_-(r_{ss},\phi_{ss})}$ in the notation of (\ref{psi0}).

In general, coherent superpositions of states of the form (\ref{psi0}) also need to be considered.  Unlike in the ordinary JCM, these superpositions do not remain approximately pure for as long as the trajectories (\ref{psit}) themselves, because of the cavity losses; for instance, in Sections 3.7--3.11 of \cite{tcburt} it is shown that superpositions of $\ket{\Psi_\pm(r_0,\phi_0)}$ with the same $\phi_0$ decohere at a rate given by $\exp(-g^2\kappa t^3/3)$.  Superpositions involving field states that start out with very different phases may decohere faster, because the photon annihilation operator, acting on the corresponding coherent states, will multiply them by different phase factors \cite{howard}.  As a result of this, one must write down the approximate steady state of the system (always neglecting spontaneous emission) as the {\em incoherent} superposition
\begin{align}
\rho_{ss} = &\frac{1}{2}\ket{\Psi_+^0(r_{ss},-\phi_{ss})}\bra{\Psi_+^0(r_{ss},-\phi_{ss})} \notag \\
&+ \frac{1}{2}\ket{\Psi_-^0(r_{ss},\phi_{ss})}\bra{\Psi_-^0(r_{ss},\phi_{ss})}
\label{rhoss}
\end{align}
As an incoherent superposition of product states, the state (\ref{rhoss}) is unentangled, or ``separable.''  

Consider, however, a {\em single} realization of the above system, which may have started from a coherent superposition of states of the form (\ref{psi0}).  Even after the system has reached a steady state, and the superposition has decohered, it may be argued (at least for as long as the individual solutions (\ref{psit}) remain approximately valid) that the decoherence is of the form of a random relative phase between the terms of the superposition, a phase that, moreover, might be knowable {\em in principle}, if we had a record of the times at which photons were emitted out of the cavity (or alternatively, through a monitoring of the transmitted field such as described in \cite{nha}).  We may then ascribe a ``conditional'' pure state to the system, of the form
\begin{equation}
\ket{\Psi_{ss}} = \frac{1}{\sqrt 2}\ket{\Psi_+^0(r_{ss},-\phi_{ss})} + \frac{e^{-i\Phi^\prime}}{\sqrt 2}\ket{\Psi_-^0(r_{ss},\phi_{ss})}
\label{psiss}
\end{equation}
where $\Phi^\prime$ is a random relative phase (time-dependent, in general, since the states in the trajectories (\ref{psit}) have overall phases that go, for short times, as $\Phi_\pm(t) = \mp g r_0 t/2$; this is analogous to the JCM and is responsible for the Rabi oscillations that occur when the two field states overlap).  For normalization purposes, it has implicitly been assumed that the two field states in (\ref{psiss}) are orthogonal, which will be approximately the case if $r_{ss}\sin\phi_{ss} \gg 1$.  For large ${\cal E}$, this condition becomes $g/2\kappa \gg1$.

Using the explicit expressions (\ref{psi0}) in (\ref{psiss}) shows that this is, in general, an entangled state, although not maximally so, since as long as $\phi_{ss} \ne 0$ the two atomic states involved, $\frac{1}{\sqrt 2}\left(e^{i\phi_{ss}}\ket e + \ket g \right)$, and $\frac{1}{\sqrt 2}\left(e^{-i\phi_{ss}}\ket e - \ket g \right)$, are not orthogonal.  Assuming the field states {\em are} orthogonal, the reduced density operator for the atom alone can be written as $\rho_A = \frac{1}{2}\ket e\bra e -\frac{i}{2}\sin\phi_{ss}\ket e\bra g +  \frac{i}{2}\sin\phi_{ss}\ket g\bra e + \frac{1}{2}\ket g\bra g$, with eigenvalues $(1\pm\sin\phi_{ss})/2$, which means that the ``entropy of entanglement,''  $E \equiv -Tr(\rho_A \log_2 \rho_A) \simeq 1- \phi_{ss}^2/2\ln 2$ for small $\phi_{ss}$.

Consider now what happens when one has a relatively small spontaneous emission rate, and a spontaneous emission event occurs.  Starting from a state like (\ref{psiss}), the atom collapses to the ground state $\ket g$, so after renormalization one has
\begin{align}
\ket{\Psi} &= \frac{1}{\sqrt 2}\ket g\ket{r_{ss}e^{i\phi_{ss}}} + \frac{e^{-i\Phi^\prime}}{\sqrt 2}\ket g\ket{r_{ss}e^{-i\phi_{ss}}} \notag \\
&= \frac{1}{2}\Bigl(\ket{\Psi_+^0(r_{ss},-\phi_{ss})}- \ket{\Psi_-^0(r_{ss},-\phi_{ss})}\Bigr) \notag \\
&\qquad + \frac{e^{-i\Phi^\prime}}{2} \Bigl(\ket{\Psi_+^0(r_{ss},\phi_{ss})}- \ket{\Psi_-^0(r_{ss},\phi_{ss})}\Bigr) 
\label{fullthing}
\end{align}
This expression has been split into two pairs of terms, the first one associated with the $u$ steady state of the field, and the second one with the $l$ steady state.  Each one of these is a superposition of the appropriate steady state ($\ket{\Psi_+^0}$ in the first parenthesis and $\ket{\Psi_-^0}$ in the second one) and another term where the field and atom have the ``wrong'' relative phase.  As shown in \cite{tcburt}, for short times, these nonstationary terms evolve by changing the phase of both field and atom at an approximate rate $\pm g/r_{ss}$, so the total time-evolved state is of the form  $\ket{\Psi(t)} = \frac{1}{\sqrt 2}(\ket{\Psi_u(t)} + {e^{-i\Phi^\prime}}\ket{\Psi_l(t)}$, with
\begin{align}
\ket{\Psi_u(t)} &= \frac{1}{\sqrt 2}\Bigl(e^{-igr_{ss}t/2}\ket{\Psi_+^0(r_{ss},-\phi_{ss})} \notag \\
&\qquad- e^{igr_{ss}t/2}\ket{\Psi_-^0(r_{ss},-\phi_{ss}+gt/r_{ss})}\Bigr) \notag \\
\ket{\Psi_l(t)} &= \frac{1}{\sqrt 2}\Bigl(e^{-igr_{ss}t/2}\ket{\Psi_+^0(r_{ss},\phi_{ss}-gt/r_{ss})} \notag \\
&\qquad- e^{igr_{ss}t/2}\ket{\Psi_-^0(r_{ss},\phi_{ss})}\Bigr).
\label{ubranch}
\end{align}
Along either one of these two ($u$ or $l$) branches, the two field states involved become approximately orthogonal as soon as $t\gg 1/g$, the JCM's ``collapse time.'' (Note that this splitting of the field into four states following a spontaneous emission event was previously discussed and illustrated in \cite{howard}; see in particular Fig.~10.8 there.)  If $r_{ss}$ is sufficiently large, the phase difference $gt/r_{ss}$ between the corresponding atomic states (along the same branch) at that time may still be quite small, in which case they will still be nearly orthogonal, and the overall state will be highly entangled.  Specifically, for either state $u$ or $l$ we find for small $gt/2r_{ss}$
\begin{equation}
E \simeq f_1(u)-f_2(u)\sin(2 gr_{ss} t) \frac{gt}{r_{ss}} - \frac{1}{8\ln 2}\left(\frac{gt}{r_{ss}}\right)^2
\label{apprent}
\end{equation}
where $u=e^{-g^2 t^2/2}$, and $f_1(u)=(u \ln((1-u)/(1+u))-\ln(1-u^2))/\ln 4$ and $f_2(u)=(2 u(1-\ln 2)+u \ln(1-u^2) + \ln((1+u)/(1-u)))/\ln 16$ are functions associated with the overlapping coherent states; at $t=0$, $f_1 = 0$ and $f_2 \simeq 0.7$, whereas after the collapse time $f_1 \to 1$ and $f_2\to 0$, and one has large entanglement provided $gt/r_{ss}$ is not too large.  Eq.~(\ref{apprent}) applies also to the superposition $ \frac{1}{\sqrt 2}(\ket{\Psi_u(t)} + {e^{-i\Phi^\prime}}\ket{\Psi_l(t)})$, regardless of the value of $\Phi^\prime$, as long as the field states in $\ket{\Psi_u(t)}$ are orthogonal to those in $\ket{\Psi_l(t)}$.

 Figure 1 shows a plot of the entropy of entanglement $E$, as a function of time, for the $u$ branch, based on the expression (\ref{ubranch}) for the system's state.  The Rabi oscillations actually cause the entanglement to peak some time before the collapse is complete.  The same figure shows also the result of a single quantum trajectory simulation (dashed line), and the result of integrating the density matrix equations of motion (dotted line), all for the same parameters and initial conditions.  The agreement between the solid and dotted lines indicates that the approximation (\ref{ubranch}) is indeed quite good. The dashed line, on the other hand, suggests that the disruption to the relative phase of the terms in (\ref{ubranch}) caused by cavity losses may sometimes reduce the entanglement obtained along an individual quantum trajectory.  
 
\begin{figure}
\includegraphics{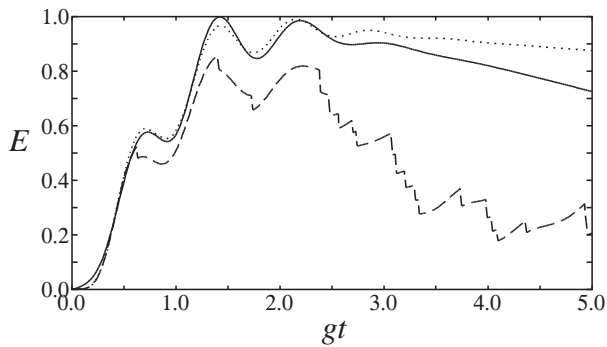}
\caption{\label{fig1} Solid line: entanglement predicted by Eq.~(\ref{ubranch}).  Dashed line: the result of a single quantum trajectory calculation. Dotted line: the result of a density matrix calculation. Model parameters: ${\cal E} = 0.7g$ and $\kappa = 0.125 g$.}
\end{figure}

The above analysis shows that, rather surprisingly, spontaneous emission may actually help generate substantial atom-field entanglement in this system, by periodically resetting the wavefunction to a state such as (\ref{fullthing}), which can later evolve into something close to a (nearly-) maximally entangled state of the form (\ref{ubranch}).  This expectation is borne out by further quantum trajectory calculations, including spontaneous emission, such as the one shown in Fig. 2.  Note the pattern: after each spontaneous emission event, the atom-field entanglement naturally goes down to zero, but then it quickly rises to, sometimes, a very high value.  This happens even if the field state at the time of the spontaneous emission event is not the steady state $\ket{r_{ss}e^{\pm i \phi_{ss}}}$, since a decomposition similar to (\ref{fullthing}) can be performed regardless of the state of the field, and the subsequent evolution will be similar to that given in (\ref{ubranch}); one only needs to use the instantaneous phase of the field (rather than the steady state value $\phi_{ss}$) in the $\ket{\Psi_\pm}$ atomic states.

\begin{figure}
\includegraphics{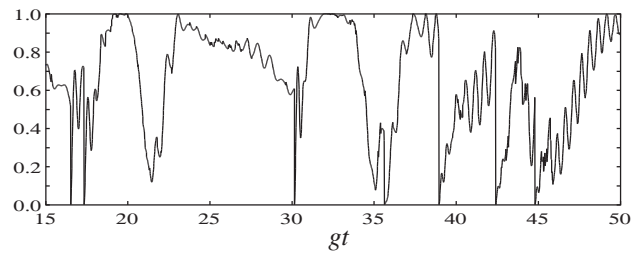}
\caption{\label{fig2}Atom-field entanglement calculated for a quantum trajectory starting from the cavity in the vacuum state and atom in the ground state. Model parameters as in Fig.~1, except that the spontaneous emission rate $\gamma = 0.4 g$.  The sharp (vertical) drops to zero entanglement correspond to spontaneous emission events, that reset the system's wavefunction to a state $\ket{\Phi}\ket g$, whatever the field state $\ket{\Phi}$ may happen to be at that instant.}
\end{figure}

When the steady state is assumed to be the mixed state (\ref{rhoss}), as opposed to the conditionally pure (\ref{psiss}), one may assume that the $u$ and $l$ branches exhibited in (\ref{fullthing}) are superimposed {\em incoherently}, although along each of them the evolution is still coherent and given by (\ref{ubranch}). After the ``collapse time,'' when all the field states involved have become orthogonal, but the phase shift $gt/r_{ss}$ is still small enough to be approximately negligible in the atomic states,  the resulting mixed state can be written schematically as $\rho = \frac 1 2  \ket{\Psi_u} \bra{\Psi_u}+ \frac 1 2  \ket{\Psi_l} \bra{\Psi_l}$, with $\ket{\Psi_u} = -\frac{1}{2} \ket 1 \left(e^{i\phi_{ss}}\ket e- \ket g\right) + \frac{1}{2} \ket 2\left(e^{i\phi_{ss}}\ket e+ \ket g\right)$ and $\ket{\Psi_l} = \frac{1}{2} \ket 3 \left(e^{-i\phi_{ss}}\ket e+ \ket g\right)  -\frac{1}{2} \ket 4\left(e^{i\phi_{ss}}\ket e- \ket g\right)$.  (Various overall phases have been absorbed in the field states $\ket 1, \ket2, \ket 3$ and $\ket 4$.)  Treating the field as a 4-dimensional system, we find, by the ``realignment criterion''  \cite{chen} that this $\rho$ still describes an entangled state. Specifically, if $G$ is the $4\times 16$ rearrangement of the $8\times 8$ matrix $\rho$, we obtain $Tr[(G G^\dagger)^{1/2}] =\sqrt 2 > 1$.  Thus, we conclude that even when the mixed nature of the steady state, for an ensemble of identically prepared systems, is considered, spontaneous emission does indeed lead to a transient entangled state, a time of the order of the collapse time after the emission event occurs.  

In both the pure and mixed-state cases, the separation of the field along each ($u$ and $l$) branch into a coherent superposition of nearly orthogonal states is essential to the generation of entanglement. This separation can be tracked by using the intensity-field correlation function \cite{pio} $h^{FT}(\tau) = \av{\sigma_+(0)a_\theta(\tau)\sigma_-(0)}/\av{\sigma_+\sigma_-}\av{a_\theta}$, where the field quadrature operator $a_\theta(\tau)$ can be calculated, at the time $\tau$, as $a_\theta(\tau) = U^\dagger(\tau)a_\theta(0)U(\tau)$, and $U(\tau)$ is the evolution operator.  Experimentally, $h^{FT}(\tau)$ gives the evolution of the transmitted field conditioned on the detection of a fluorescent photon (i.e., a spontaneous emission event) at the time $\tau = 0$.  The approximation (\ref{ubranch}) can then be used to calculate it; the result (which does not depend on whether $\ket{\Psi_u}$ and $\ket{\Psi_l}$ are added coherently or incoherently, as long as the field states in $\ket{\Psi_u}$ are orthogonal to those in $\ket{\Psi_l}$) is plotted in Fig.~3 (solid line), where the similarity to the pure-state entanglement curve (Fig.~(2)) is readily apparent. To better understand this similarity, one may consider the following approximate result for small $gt/r_{ss}$:
\begin{equation}
h^{FT}-1\simeq (\tan\phi_{ss} + u \sin(2 g rt))\frac{gt}{2 r_{ss}} -\frac{1}{4}\left(\frac{gt}{r_{ss}}\right)^2
\label{hft}
\end{equation}
with $u=\exp(-g^2 t^2/2)$, as before. Like Eq.~(\ref{apprent}), Eq.~(\ref{hft}) shows Rabi oscillations (although with the opposite sign) that die away at the collapse time. In both cases, the initial rise of the curves is due to the growing separation, in phase space, of the two field states making up the $u$ or $l$ branch, although the entanglement eventually saturates, around the collapse time, whereas $h^{FT}-1$ may continue to grow (due to the term $\tan\phi_{ss}gt/2 r_{ss}$) up to a time of the order of $1/2\kappa$. 

\begin{figure}
\includegraphics{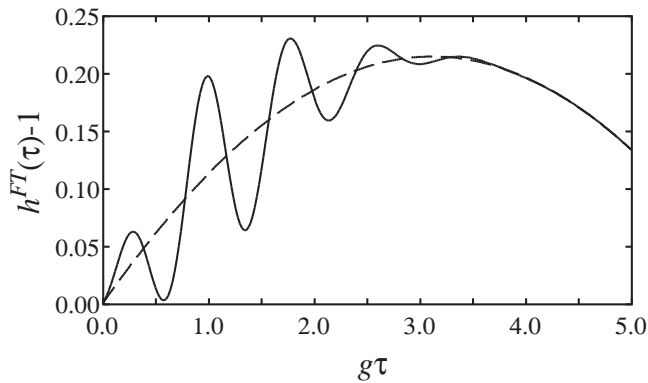}
\caption{\label{fig3} Solid line: $h^{FT}(\tau)-1)$ for the same parameters as in Fig.~1, based on the expressions (\ref{ubranch}).  Dashed line: the result if the pure state $\ket{\Psi_u(t)}$ (or $\ket{\Psi_u(t)}$) is replaced by an incoherent superposition of $\ket{\Psi_+}$ and $\ket{\Psi_-}$.}
\end{figure}

We conclude that, through the collapse time, the correlation function $h^{FT}(\tau)$ may be used to track the physical processes underlying the growth of the atom-field entanglement in the system, subsequent to a spontaneous emission event, although, in order to make the entanglement correspondence quantitative, a fair amount of theory needs to be assumed.  In particular, note that it is not enough to observe an increase in $h^{FT}-1$ to conclude that entanglement must be growing, since $h^{FT}-1$ would rise, as a result of the separation of the field states, even if the superposition of states making up $\ket{\Psi_{u,l}(t)}$ in Eq.~(\ref{ubranch}) was completely incoherent (dashed line in Fig.~(3); or set $u=0$ in Eq.~(\ref{hft})), in which case there would be no entanglement at all. The Rabi oscillations are thus critical evidence that the superposition is coherent and the underlying state is entangled.  Ironically, these oscillations disappear around the collapse time, just when one expects entanglement to be largest.  In an experimental setting, the underlying coherence might be revealed using methods such as those suggested in \cite{auffeves}, to reverse the sign of rotation of the field states and bring back the oscillations. 

The feasibility of exploring this entanglement phenomenon is within
reach of current strong-coupled optical cavity QED experiments. It will
open the strong driving parameter space that is different from the one
most explored to date, the weak driving regime. Further work is
necessary to expand this to the case of more than one atom \cite{raizen}, and to properly account for multiple (partly overlapping) spontaneous emission events.

We are grateful to H. J. Carmichael for very helpful discussions.  This work has been supported by the NSF.

\end{document}